\documentstyle[12pt]{article}

\def\be{\begin{equation}}
\def\ee{\end{equation}}
\def\ba{\begin{array}{c}}
\def\ea{\end{array}}

\def\ben{$$}
\def\een{$$}
\begin{document}

\titlepage

\begin{center}{\Large \bf
Pseudo-Hermitian version of the charged harmonic oscillator and
its ``forgotten" exact solutions
 }\end{center}

\vspace{5mm}

\begin{center}

Miloslav Znojil

\vspace{3mm}

Nuclear Physics Institute of Academy of Sciences of the Czech
Republic,

 250 68 \v{R}e\v{z}, Czech Republic

e-mail: znojil@ujf.cas.cz


\vspace{5mm}

\end{center}

\vspace{5mm}

\vspace{5mm}

\section*{Abstract}

An unusual type of the exact solvability is reported. It is
exemplified by the Coulomb plus harmonic oscillator in $D$
dimensions after a complexification of its Hamiltonian which
keeps the energies real. Infinitely many bound states are found
in closed form which generalizes the popular harmonic-oscillator
states at zero charge and even parity.  Apparently, the model is
halfway between exact and quasi-exact.

\vspace{9mm}

\noindent
 PACS 03.65.Ge,
03.65.Fd



\newpage

\section{Introduction}

Schr\"{o}dinger equation for the shifted and charged harmonic
oscillator in $D$ dimensions reads
 \be
\left[-\,\frac{d^2}{dr^2}+ \frac{\ell(\ell+1)}{r^2}+ \frac{f}{r} +
2\,g\,r + r^2\right]\,
 \varphi(r) = E\, \varphi(r), \ \ \ \ \ r \in (0, \infty).
 \label{SEH}
 \ee
Thirty years ago, Andr\'{e} Hautot \cite{Hautot} noticed that at
certain non-vanishing couplings $f$ and/or $g$ it may posses
elementary solutions for the equidistant set of the
energy levels
 \be
 E=
 E^{(Hautot)}_{n,\ell} = 2n+2+{\cal L} - g^2
 ,
 \label{Hautots}
 \ee
 \ben
 n = 0, 1, \ldots \, ,\ \ \ \ \ \
 {\cal L}= {\cal L}(\ell)=2\ell+1 = D-2, D, D+2,\ldots
 \een
which do not depend on the charge $f$. The charge
itself is not arbitrary (for this reason, the models of this
type are called quasi-exactly solvable, QES). One has to
evaluate the admissible values of $f=f_n$ as real roots of
certain polynomials of the $(n-1)-$st degree (see below).

The undeniable mathematical as well as physical appeal of QES
solutions has been revealed by many independent authors whose
work was summarized by Alex Ushveridze \cite{Ushveridze}. Very
recently, the QES models helped to clarify some counterintuitive
formal features of the so called ${\cal PT}$ symmetric quantum
mechanics of Bender et al \cite{BB} who replaced the usual
Hermiticity $H = H^\dagger$ by the mere commutativity of the
Hamiltonian with the product of parity ${\cal P}$ and time
reversal ${\cal T}$. In the early stages of development, the
studies of this formalism were strongly motivated by its
relevance in field theory \cite{BM}. In such a setting, it was
very impressive when Bender and Boettcher \cite{BBjpa}
demonstrated that in contrast to the current Hermitian case,
quartic polynomial oscillators belong to the QES class after
their appropriate ${\cal PT}$ symmetrization (cf. also ref.
\cite{quartic} for more details).

The charged harmonic oscillator (\ref{SEH}) does not possess
similar appeal in field theory but it was still amusing to reveal
in paper \cite{I} that its Schr\"{o}dinger equation does not lose
its partial elementary solvability even after the weakening of the
Hermiticity to the mere ${\cal PT}$ symmetry of its Hamiltonian.
In accord with the Buslaev's and Grecchi's recipe \cite{BG} we
used the shifted coordinates $ x \in (-\infty, \infty)$ on complex
line $r(x) = x - i\,\varepsilon$ at a constant distance
$\varepsilon> 0$ from the real axis,
 \be
 \left[-\,\frac{d^2}{dx^2}+
 \frac{\ell(\ell+1)}{r^2(x)}+ i\,\frac{F_n}{r(x)} +
 2\,i\,b\,r(x) +  r^2(x)\right]\,
 \psi_n(x) = E_n\, \psi_n(x).
 \label{SE}
 \ee
Unfortunately, we only analyzed the $\ell = 0$ solutions in the
quasi-odd regime (cf. a more detailed explanation below).

Since the publication of paper \cite{I} a significant progress has
been achieved in the interpretation of the non-Hermitian
equations. Several authors \cite{AM} emphasized
that within the domain of quantum mechanics, the ${\cal
PT}$ symmetry of eq.~(\ref{SE}) should be replaced by the
(formally equivalent but mathematically more natural)
requirement of the pseudo-Hermiticity of the Hamiltonian,
 \be
 H = \eta^{-1} H^\dagger \eta, \ \ \ \ \ \ \ \eta = \eta^\dagger.
 \label{pseudo}
 \ee
One of the oldest illustrations of the efficiency of the use of
the pseudo-Hermitian Hamiltonians (\ref{pseudo}) with an auxiliary
indefinite metric $\eta$ appears in the relativistic quantum
mechanics where it very naturally originates from the Feshbach's
and Villars' Hamiltonian formulation of the Klein Gordon equation
for the particles with zero spin \cite{Constantinescu}. The same
choice of a suitable invertible $\eta$ helps to clarify some
contemporary problems in the cosmological models based on the
equations of Bryce de Witt~\cite{Ernie}.

In the light of eq. (\ref{pseudo}), physical interpretation of
the non-Hermitian bound states is more transparent and does not
depend too much on the specification of the operator $\eta$
itself. This operator only plays the role of a certain auxiliary
transformation of the dual Hilbert space. For a detailed
explanation of this idea we recommend the older review
\cite{Geyer} where the physical meaning of the nontrivial
``metric" $\eta$ was illustrated by its emergence in the
many-fermion models where $\eta \neq I$ characterizes the so
called Dyson's mapping of the ``physical" (and Hermitian)
fermionic Hamiltonians onto their ``more easily solvable"
$\eta-$Hermitian bosonic equivalents~\cite{Jolos}.

One must keep in mind the non-uniqueness of the metric $\eta$
which belongs to the given $H$. According to A. Mostafazadeh
\cite{AMb} one can replace the initial indefinite (and, in
particular, ${\cal PT}-$symmetric) metric $\eta_1$ by an
alternative $\eta_2$ {\em which is positive definite}. In the
other words, the puzzling quasi-unitary evolution generated by
the indefinite $\eta_1$ \cite{quasi} may be declared an artifact
of our constructions. {\it Vice versa}, all the phenomenological
considerations should necessarily be related to the positive
definite version $\eta_2>0$ of the metric (one can show that is
exists for all the diagonalizable Hamiltonians \cite{AM,AMb}).
Then, the time evolution remains compatible with the
probabilistic interpretation of the norm in the Hilbert space of
states.

In the light of the possible peaceful coexistence of $\eta_1={\cal
P}$ with some $\eta_2>0$ in eq.~(\ref{pseudo}), our attention has
been re-directed to the solutions of eq. (\ref{SE}) which diverge
in the simple-minded Hermitian limit $\varepsilon \to 0$ and which
were omitted from the scope of our preceding study \cite{I}.  We
are going to correct the omission now. Our expository section
\ref{2} will outline an improvement of the method which generates
the old QES solutions of ref. \cite{I}. Section \ref{3} will then
modify the basic ansatz which opens the way towards the new (or,
in the wording of our title, ``forgotten") QES solutions of
eq.~(\ref{SE}). Some of their properties and possible applications
will be finally discussed in section~\ref{4}.

\section{The standard quasi-exact solutions \label{2}}

As long as the differential eq.~(\ref{SE}) is of the second order,
its general solution is a superposition of some two linearly
independent solutions. This independence may be deduced from
their available leading-order form near the origin,
 \be
 \psi_n(r) = c_-\psi_{n}^{(-)}(r)+c_+\psi_{n}^{(+)}(r), \ \ \ \
 \
 \label{spolu}
\ \ \ \ \ \ \
 \psi_{n}^{(\pm)}(r) = {\cal O} \left ( r^{1/2\mp (\ell+1/2)}
 \right )
 \ .
 \ee
In the spirit of ref. \cite{I} one usually searches for the
polynomial solutions compatible with the correct physical boundary
conditions in the origin (i.e., $c_+ = 0$ \cite{comment}) as well
as with their asymptotic normalizability. In the spirit of the
general QES philosophy one may meet both these requirements by
employing the special elementary ansatz of the very common
harmonic-oscillator-like form
\be
\psi(x)=\psi_{n}^{(-)}(x) = e^{-r^2/2-i\,b\,r}\ \sum_{n=0}^{N}\
h_n^{(-)} (i\,r)^{n+\ell+1}\  , \ \ \ \ \ r=r(x) = x -
i\,\varepsilon
  \label{ransatz}.
 \ee
The construction of the solutions of this type degenerates to the
insertion of eq. (\ref{ransatz}) into the differential eq.
(\ref{SE}) which gives the homogeneous set of
$N+2$ linear algebraic equations for the $N+1$ coefficients
$h_j^{(-)}$. We may drop the superscript and turn to
the explicit non-square matrix form of the latter equations,
\be
 \left( \begin{array}{ccccc} B_{0} & C_0&  & &  \\
A_1&B_{1} & C_1&    & \\ &\ddots&\ddots&\ddots&\\
&&A_{N-1}&B_{N-1}&C_{N-1}\\ &&&A_{N}&B_{N}\\ &&&&A_{N+1}
\end{array} \right) \left ( \ba h_0\\ h_2\\ \vdots \\ h_N \ea \right)
= 0\
 \label{recu}
 \ee
with elements
 \ben
\ba
 A_n =A_n^{(-)} =b^2+ 2n  +{\cal L}-E, \ \ \ \ \ \ \
 B_n =  B_n^{(-)} =-(2n+1+{\cal L})b-F,\\
 C_n =  C_n^{(-)} =(n+1)\,(n+1+{\cal L}),\ \ \ \ \ \ \
  {\cal L} = 2\ell+1, \ \  \
  \ \ \ \ n = 0, 1, \ldots \ .
  \ea
  \label{eleme2}
   \een
This is a finite-dimensional and over-determined linear algebraic
re-incarnation of the original differential equation (\ref{SE}).
Its matrix structure enables us to define the wave function (i.e.,
all its energy-dependent coefficients) as determinants,
\be
 h_{j-1}= \frac{h_N}{(-A_j)(-A_{j+1}) \ldots (-A_{N})}
 \cdot \,\det
 \left( \begin{array}{ccccc} B_{j} & C_j&  & &  \\
A_{j+1}&B_{j+1} & C_{j+1}&    & \\ &\ddots&\ddots&\ddots&\\
&&A_{N-1}&B_{N-1}&C_{N-1}\\ &&&A_{N}&B_{N}
\end{array} \right)
 \label{exrecu}
 \ee
with $j = N, N-1, \ldots, 1$ and under any choice of the
normalization $h_N\neq 0$.

The latter normalization convention converts the last row
$A_{N+1}h_N=0$ of eq. (\ref{recu}) into the constraint
 \be
 E=E^{(-)}= 2N+2 +{\cal L} +b^2.
 \label{energoo}
 \ee
In the other words, the condition of the mutual compatibility of
the original over-determined linear system~(\ref{recu}) fixes the
energy which coincides with the old Hautot's formula
(\ref{Hautots}). At any $N= 0, 1, \ldots$ the energy is
an increasing function of the angular momentum $\ell$ or ${\cal
L}$ and of the size of the shift~$b$. The QES construction is
complete and
\begin{itemize}

\item
simplifies the prescription of ref. \cite{I} (where the
special cases of eqs. (\ref{recu}) and (\ref{exrecu}) contained the
less compact matrix with four diagonals),

\item
leads to the polynomial wave functions (with the closed form
(\ref{exrecu}) of the coefficients),

\item
preserves the Hautot's explicit form (\ref{energ}) of the
energies,

\item
reduces the differential Schr\"{o}dinger equation to its $n-$
dimensional square-matrix form.

\end{itemize}

 \noindent
The first observation (simplification) is a marginal technical
merit due to our transition to a better ansatz. In contrast, the
last feature of the QES solutions remains highly unpleasant as
it forces us to guarantee that the related secular determinant
vanishes,
 \be
 \det
 \left( \begin{array}{ccccc} B_{0}^{(-)}
  & C_{0}^{(-)}
 &  & &  \\
 A_{1}^{(-)}&B_{1}^{(-)} & C_{1}^{(-)}&
 &
 \\ &\ddots&\ddots&\ddots&\\ &&A_{N-1}^{(+)}
 &B_{N-1}^{(-)}&C_{N-1}^{(-)}\\ &&&A_{N}^{(-)}&B_{N}^{(-)}
 \end{array}
  \right)= 0\ .
 \label{qesc}
 \ee
Such a constraint determines the set of the  $N+1$ admissible
couplings $F = F_k^{(-)}(N)$ \cite{Hautot} and its purely
numerical nature is an example of the most serious practical
shortcoming of the majority of the QES models~\cite{Ushveridze}.
We are now going to describe a remarkable exception from this
discouraging rule.

\section{Nonstandard, quasi-even QES states  \label{3}}

\subsection{The concept of quasi-parity \label{3.1}}

Above we emphasized that at a fixed, non-vanishing shift
$\varepsilon>0$ the ambiguity of the metric $\eta$ opens the
possibility of using the ``simpler"
conjugation (\ref{pseudo}) with $\eta=\eta_1= {\cal P}$ during the
explicit constructions of the solutions while switching
to their ``physical" re-interpretation based on
an alternative scalar product
with the positive definite metric~$ \eta_2>0$.
In such an approach
 both components $c_- \neq 0$ and $c_+ \neq 0$ of
wave functions in eq. (\ref{spolu}) may be equally useful.

Once we relax the redundant boundary conditions in the origin we
get more solutions of course. One of the most transparent
illustrations of the emergence of the additional
``quasi-even" $c_+ \neq 0$ solutions was described in ref.
\cite{ptho} where eq.~(\ref{SE}) has been solved at the
vanishing $b = F_n=0$. The superscripts in the resulting states
$\psi_{n}^{(+)}(r)$ and $\psi_{n}^{(-)}(r)$ have been
interpreted as the so called quasi-parity. The introduction of
this concept was motivated by the observation that the
quasi-parity degenerated to the current parity at $b = F_n=\ell
= 0$ (cf. eq.~(\ref{spolu})).

Its independent additional support appeared in ref.~\cite{Quesne}
revealing its connection with the ${\cal PT}$ parity and with the
pseudo-norm using $\eta=\eta_1= {\cal P}$. It even plays its role
in the supersymmetric quantum mechanics (cf. ref.~\cite{ptsusy}
for more details) but is missing from our ansatz (\ref{ransatz})
inherited from ref. \cite{I}. As we understand it now, ansatz
(\ref{ransatz}) is unnecessarily restrictive as it represents
merely quasi-odd solutions. From its generalization
\be
\psi_{n}(r) = e^{-r^2/2-i\,b\,r}\ \sum_{n=0}^{N}\ h_n
(i\,r)^{n-\ell}\   \label{ansatz}
 \ee
(where we dropped all the superscripts for the time being) one
can always return to the old quasi-odd option via the additional
$^{(-)}-$superscripted constraint
 \be
h_0^{(-)}=h_1^{(-)}=\ldots = h_{{\cal L}-1}^{(-)}=0\ .
\label{condi}
 \ee
Whenever necessary, the $^{(+)}-$superscripted ``quasi-even" QES
solutions may be characterized by the alternative criterion
\be
|h_0^{(+)}| + |h_1^{(+)}| +\ldots + | h_{{\cal L}-1}^{(+)}| > 0.
\label{condibe}
 \ee
We are now close to our key claim that the structure of the
quasi-even QES solutions is exceptionally simple.

\subsection{QES states having the even quasi-parity \label{3.2}}

The source of the latter claim lies in the improved ansatz
(\ref{ansatz}) which leads to the same equation (\ref{recu}) with
the new matrix elements
 \ben
\ba
 A_n =A_n^{(+)} =b^2+ 2n  -{\cal L}-E, \ \ \ \ \ \ \
 B_n =  B_n^{(+)} =-(2n+1-{\cal L})b-F,\\
 C_n =  C_n^{(+)} =(n+1)\,(n+1-{\cal L}),\ \ \ \ \ \ \
  {\cal L} = 2\ell+1, \ \  \
  \ \ \ \ n = 0, 1, \ldots \ .
  \ea
  \label{elemeff2}
   \een
The energy formula (\ref{energoo}) is only marginally modified,
 \be
 E=E^{(+)}= 2N+2 -{\cal L} +b^2.
 \label{energ}
 \ee
Still, we immediately notice the much more important difference
connected with the presence of the vanishing matrix element $
C_{{\cal L}-1}^{(+)} =0$ in the upper diagonal of our new form of
the QES secular equation. This means that the
$^{(+)}-$superscripted secular determinant may be re-written as
the product of a ``small", ${\cal L}-$dimensional
 \ben
  {\cal S}^{(S)} =\det
 \left( \begin{array}{ccccc} B_{0}^{(+)}
  & C_{0}^{(+)}
 &  & &  \\
 A_{1}^{(+)}&B_{1}^{(+)} & C_{1}^{(+)}&
 &
 \\ &\ddots&\ddots&\ddots&\\ &&A_{{\cal L}-2}^{(+)}
 &B_{{\cal L}-2}^{(+)}&C_{{\cal L}-2}^{(+)}\\
  &&&A_{{\cal L}-1}^{(+)}&B_{{\cal L}-1}^{(+)}
 \end{array}
  \right)\
 \label{curare}
 \een
with another determinant ${\cal S}^{(L)}$ of a ``large", $(N-{\cal
L})-$dimensional matrix. The latter factor
 \ben
  {\cal S}^{(L)} =\det
 \left( \begin{array}{ccccc} B_{{\cal L}}^{(+)}
  & C_{{\cal L}}^{(+)}
 &  & &  \\
 A_{{\cal L}+1}^{(+)}&B_{{\cal L}+1}^{(+)} & C_{{\cal L}+1}^{(+)}&
 &
 \\ &\ddots&\ddots&\ddots&\\ &&A_{N-1}^{(+)}
 &B_{N-1}^{(+)}&C_{N-1}^{(+)}\\ &&&A_{N}^{(+)}&B_{N}^{(+)}
 \end{array}
  \right)\
 \label{qesrec}
 \een
precisely coincides with the left-hand side expression in eq.
(\ref{qesc}) which guarantees, in its turn, the existence of the
QES solutions with the property (\ref{condi}). The use of the
condition  ${\cal S}^{(L)}=0$ would return us back to the old
quasi-odd ansatzs of section \ref{2}. In what follows we
shall ignore these solutions as standard and assume that ${\cal
S}^{(L)} \neq 0$.

\subsection{Facilitated QES constructions}

After one concentrates attention solely to the quasi-even states,
the QES construction degenerates, basically, to the secular
equation ${\cal S}^{(S)}=0$. One verifies easily that the
acceptance of this condition is consistent with the quasi-parity
(\ref{condibe}). The wave function coefficients themselves remain
formally determined by the $^{(+)}-$superscripted version of the
determinants (\ref{exrecu}) whose dimension grows with $N$. For
this reason we recommend a re-interpretation of these coefficients
as quantities evaluated by the recurrences initiated at an initial
$h_N^{(+)} \neq 0$ and defining $h_{N-1}^{(+)}$, $h_{N-2}^{(+)}.
\ldots$ step by step,
 \be
 \left( \begin{array}{ccccc}
-2N&\beta_1-F & 4-2{\cal L}&    & \\ &\ddots&\ddots&\ddots&\\
&&-4&\beta_{N-1}-F&N^2-N\,{\cal L}\\ &&&-2&\beta_N-F
\end{array} \right)
 \left ( \ba h_0^{(+)}\\ \vdots \\ h_{N-1}^{(+)}
  \\ h_N^{(+)} \ea \right)
= 0\ .
 \label{recupe}
 \ee
We abbreviated here $\beta_n \equiv -(2n+1-{\cal L})\,b$ and note
that in the last step of these recurrences, the secular equation $
{\cal S}^{(S)} =0$ replaces the redundant definition of a ghost
coefficient $h_{-1}^{(+)} = 0$. Finally, the explicit form of our
secular equation
 \be
 \det
 \left( \begin{array}{ccccc} \beta_0-F & 1-{\cal L}&  & &  \\
-2N&\beta_1-F & 4-2{\cal L}&    & \\ &\ddots&\ddots&\ddots&\\
&&-2(N+3-{\cal L})&\beta_{{\cal L}-2}-F&1-{\cal L}\\
&&&-2(N+2-{\cal L})&\beta_{{\cal L}-1}-F
\end{array} \right)
 =0 \
 \label{recuperare}
 \ee
specifies the family of the admissible charges $F_N =
F_k^{(+)}(N)$ in a way which remains virtually purely
non-numerical for the first few dimensions ${\cal L}$.

\subsubsection{${\cal L} = 2$}

One of the key reasons why both the formal appeal and practical
importance of the quasi-even spectrum (\ref{energ}) remained
unnoticed in ref. \cite{I} was purely psychological. Indeed, at
the simplest choice of ${\cal L} = 1$ in eq. (\ref{recuperare})
(which may mean {\em both} the $s-$wave in three dimensions {\em
and} an even state at $D=0$) one does not obtain anything new.
Equation (\ref{recuperare}) provides the single root
$F_0^{(+)}(N)= 0$ and one just reveals the well known fact that at
the vanishing eigencharge our model degenerates to the linear
harmonic oscillator defined on the whole line.

Let us move, therefore, to the first nontrivial choice of ${\cal
L}=2$ corresponding to the $p-$wave in two dimensions or to the
$s-$wave in four dimensions. This gives the following two series
of the fully non-numerical eigencharges,
\be
 F_{[1,2]}(N)=\pm \sqrt{\left( b^{2}+2N\right)}\label{refex}.
 \ee
The wave functions retain the even quasi-parity in a
way compatible with eq. (\ref{condibe}),
 \ben
 h_{0\,[1,2]}^{(+)}(N) = -\frac{1}{2N}\,
  \left [ {b+F_{[1,2]}(N)} \right
 ] \,h_{1\,[1,2]}^{(+)}(N) .
 \een
Both the eigencharges grow with the increasing size of the shift
$b$ and with the number $N$ of nodes (i.e., with the
energy~$E_N$). In the coupling-energy plane our QES states may be
visualized as families located along certain curved lines which,
in a way, lie somewhere in between the harmonic-oscillator $F=0$
straight line and the numerous Hautot's sets of roots $F_k(N)$
each of which is defined at a fixed energy or integer $N$.
Sometimes, the similar families of the bound states with the
``energy = constant" property are being called Sturmians. In this
sense one could speak here about a certain further generalization
of the latter concept.

\subsubsection{${\cal L} = 3$}

A mild formal shortcoming of the present Coulomb + harmonic
oscillators lies in a quick growth of complexity of eq.
(\ref{recuperare}) for the larger ${\cal L}$. At any  ${\cal L}
\geq 3$ one should not be tempted to generate the formulae
verifying, say, the smooth $N-$ and $b-$dependence of the
eigencharges.  In practice, the other approaches may definitely
prove preferable.

Even at the very first ${\cal L}=3$ the comparatively compact form
of our eq. (\ref{recuperare}),
\be
  \det
\left (
\begin{array}
[c]{ccc}%
2b-F & -2 & 0\\ -2N & -F & -2\\ 0 & -2N+2 & -2b-F
\end{array}
\right ) = 0,
 \label{ouk}
 \ee
should not inspire a search for the triplet of charges $\{F_0,
F_1,F_2\}$ via the closed (i.e., Cardano) formulae (the reader is
recommended to try to generate them using the computer symbolic
manipulations in order to see that they really are enormously
clumsy).  A significantly better strategy consists in an
elimination of the (unique) value of $N$ from the above secular
determinant (\ref{ouk}) giving
\[
 N=N(F,b)=-\frac{1}{8F} \left({4Fb^{2}+8b-F^{3}-4F}\right )\ .
 \]
After we fix any left-hand-side integer $N$ we may pick up the two
eigenshifts $b_{[1,2]}$ as elementary functions of the
indeterminate variable~$F$ (we skip the details which are
trivial).

\subsubsection{${\cal L} > 3$}

The main advantage of the semi-implicit techniques of the solution
of eq.~(\ref{recuperare}) is that they may work at a few larger
integers ${\cal L}>3$. For illustration, the choice of ${\cal
L}=5$ (which corresponds to the $d-$wave in three dimensions)
leads already to the purely numerical determination of
eigencharges $F$.  In an alternative approach we eliminate
\[ N^\pm=
\frac{1}{512F}\left [ -768b-256Fb^{2}+768F+40F^{3}\pm
24\sqrt{\left( 1024b^{2}+192bF^{3}+512F^{2}+F^{6}\right) }
\right ]
\]
and recommend the graphical determination of the eigenvalues $F=F(b)$
and/or $b=b(F)$ afterwards.

The most practical possibility consists in a direct selection of a
suitable shift~$b$ followed by the subsequent diagonalization of
the purely numerical matrix. In an illustration using  ${\cal
L}=3$ and $b=5$ one gets the three eigencharges
\[
F = \{10.757,\ -10.400, \ -0.35755 \}
 \]
at the smallest possible $N=2$. In an opposite-extreme using
$N=1000$ the computed values
 \[
 F = \{89.98,\ -89.975, \
-0.0049407 \}
 \]
already lie very close to their large$-N$ estimate obtainable in
closed form,
\[ F \approx
\{\sqrt{8N},\ -\sqrt{8N}, \ -b/N \} \approx \{89.44,\ -89.44, \
-0.005   \}.
 \]
This simplifies the verification of the reality of the
eigencharges and confirms the smoothness of their
$N-$dependence.  Such a type of calculation is very quick and
gives results sampled at $ {\cal L}=4$ and $b=5$ in Table~1.

\section{Discussion \label{4}}

The main merit of our key eq. (\ref{recuperare}) (which defines
the eigencharges) is that its dimension is independent of the
quantum number~$N$. Equation (\ref{refex}) is the best
illustration of the related new form of the solvability which we
intended to describe here. Still, the principle of the whole
construction is more general and one might try to apply the
similar recipe to the quartic oscillator of refs.
\cite{BBjpa,quartic}, to the sextic oscillators studied by many
authors \cite{sextic}, to the decadic oscillator of ref.
\cite{decadic} and to the numerous existing
modifications~\cite{Ushveridze} of these most popular or
``canonical" models.

\subsection{Non-orthogonal QES states as a basis?}

In our particular example, the strength of the Coulombic
interaction appears to be an energy- or $N-$dependent quantity.
One deals with an ${\cal L}\geq 2$ generalization of the common
${\cal L}=1$ harmonic oscillator. Its most important features are
an apparent completeness ``in a relevant subspace" (a guess
inspired by their infinite number) and a compact form (reflecting
the $N-$independent evaluation of the eligible charges, each of
which is selected as a function of the main quantum number $N=0,
1, \ldots$).  Both these features make our infinite set(s) of the
quasi-even states very similar to the ordinary harmonic oscillator
basis.  In the context of concluding remarks, let us pay some
attention to the possible analogies of the latter type.

Firstly, due to the non-Hermiticity of the Hamiltonian $H(F) =
H(0)+ F\,W$ we have to distinguish between the left (= double-ket)
and right (= single-ket) QES eigenstates,
 \be
 \left [ H(0) + F_N\,W
 \right ]\ |N\rangle = E_N\ |N\rangle,
 \label{one}
 \ee
 \be
 \langle \langle N | \
 \left [ H(0) + F_N\,W
 \right ] = E_N\ \langle \langle N |
 \label{two}.
 \ee
The integer $N$ numbers the energies $E_N$ as well as the selected
charges $F_N= F_[k]^{(+)}(N)$ so that the left and right
eigenstates exist at the common energies (\ref{energ}) and charges
[say, (\ref{refex})]. The wave functions are defined in closed
form, as polynomials in the coordinates (cf.~(\ref{ansatz})) and
in the couplings and quantum numbers (cf.~(\ref{exrecu})). This is
of paramount importance, making our quasi-even QES solutions
extremely similar to the even bound states of the exactly solvable
chargeless oscillators which form one of the most popular complete
bases in $L_2(0,\infty)$.

In a tentative parallel one could search for the appropriately
weakened biorthogonality relations.  This is a real mathematical
challenge since our QES solutions are only defined at the
exceptional and $\ell-$ and $N-$dependent charges. Still, one can
easily verify the manifest non-orthogonality of the pairs of many
randomly selected QES states. Of course, the closed form as well
as the infinite number of these states inspires their use in a
perturbative or variational context.

For the similar purposes one has to truncate their set to a
finite subset (${\cal N} < \infty$), assuming that the
matrix of their overlaps
 \ben
 Q_{m,n} = \langle \langle m|n\rangle, \ \ \ \ \ \ \
 m,n = 0, 1, \ldots
 {\cal N}
 \een
is invertible. We may then proceed, say, in the variational spirit and
reduce our Hilbert space and its dual to the finite-dimensional
subspaces spanned by our subset of the selected QES
eigenvectors. The approximate identity operator becomes
defined by the usual series
 \ben
 I = \sum_{m,n=0}^{\cal N}\ | m \rangle\,R_{m,n}\,\langle\langle n |
 \een
where $R=Q^{-1}$ is, in general, fully non-diagonal. Also the
Hamiltonian $H(F)$ itself becomes approximated by a non-diagonal
matrix. At almost all $F$, the search for the energies $E=E(F)$
becomes, therefore, a numerical task.

\subsection{Speculations about applicability}

In spite of the unpleasant character of the latter conclusion, one
should still feel the difference between a fully general matrix
diagonalization and our ``next-to-solvable" Coulomb + harmonic
problem {\em considered at any charge } $F$,
 \be
 \left [ H(0) + F\,W
 \right ]\ |\Psi \rangle = E(F) \ |\Psi \rangle\ .
 \label{obone}
 \ee
In particular, our Schr\"{o}dinger equation may be (e.g.,
perturbatively) connected to its special cases with QES character.
In an attempted step towards making such a connection explicit,
let us imagine that equations (\ref{one}) and (\ref{two}) share
the energy and charge (though not the eigevectors) at every given
$N$. This means that, respectively, we have the relations
 \be
 \langle \langle N | \
 \left [ H(0) + F_M\,W
 \right ]\ |M\rangle = E_M\  \langle \langle N |M\rangle,
 \label{oned}
 \ee
 \be
 \langle \langle N | \
 \left [ H(0) + F_N\,W
 \right ]\ |M\rangle = E_N\ \langle \langle N  |M\rangle
 \label{twod}
 \ee
the subtraction of which gives the strongest constraint
 \be
 \left (
 F_M-F_N
 \right )\,
 \langle \langle N |\, W\,
  |M\rangle =
\left (
 E_M-E_N
 \right )\,
   Q_{N,M}.
 \label{biogg}
 \ee
This relation is an immediate generalization of the
bi-orthogonality of the states which would result from it in the
hypothetical case of the subscript-independent charges $
F_M=F_N$.

Let us now return to the relations (\ref{oned}) and (\ref{twod})
and deduce the matrix form of the Coulomb + harmonic Hamiltonian
$H(F)$ at any value of the charge,
 \be
 \langle \langle N | \
 H(F)
\ |M\rangle = \left ( F-F_M \right )\,
 \langle \langle N | \
W \ |M\rangle +
 E_M\    Q_{N,M}\ .
 \label{tonedr}
 \ee
This formula will help us to avoid the tedious and most time
consuming part of the diagonalization of $H(F)$, viz, the
evaluation of the input matrix elements which precedes the
solution of the problem (\ref{obone}) studied in its explicit
matrix form which defines, in principle, all the necessary
components $p_N = \langle \langle N|\Psi\rangle$ of the wave
functions
 \ben
|\Psi\rangle =
\sum_{m,n=0}^{\cal N}\ | m \rangle\,R_{m,n}\,p_n.
 \een
At this moment we have to re-emphasize that we do not intend to
perform any numerical calculations. We rather wish to stress the
helpful role which can be played by the QES states. In such a
context, the number of the necessary input matrix elements
encounters the most drastic reduction after the insertion of eq.
(\ref{tonedr}) in eq.~(\ref{obone}),
 \be
 \sum_{K,J}\,
 \left (
 F-F_N
 \right )\,
 \langle \langle N |\, W\,
  |K\rangle\,R_{K,J}\ p_J
  =
\left (
 E-E_N
 \right )\ p_N.
 \label{SEF}
 \ee
Next, the necessary input information is further reduced by the
generalized biorthogonality relations (\ref{biogg}) which express
all the off-diagonal elements of the (in our example, Coulombic)
operator $W$ in terms of the known overlap matrix $Q$. In this way
the numerical or perturbative diagonalization of the matrix
Schr\"{o}dinger eq.~(\ref{SEF}),
 \be
 \left (
 F-F_N
 \right )\,
 \sum_J
 \left [
 T_N\,R_{N,J}+
 \sum_{K\neq N}\,
  \frac{E_N-E_K}{F_N-F_K}\
 Q_{N,K}\,R_{K,J}
 \right ]
  \ p_J
  =
\left (
 E-E_N
 \right )\ p_N,
 \label{SET}
 \ee
 \ben
 \ \ \ \ \ \ \ \  \  \ \ \ \ \ \
 \ \ \ \ \ \ \ \  \  \ \ \ \ \ \
 \ \ \ \ \ \ \ \  \  \ \ \ \ \ \
 \ \ \ \ \ \ \ \  \  \ \ \ \ \ \
 N = 0, 1, \ldots, {\cal N}
 \een
will require the independent input evaluation of the mere diagonal
matrix elements $T_N=\langle \langle N |\, W\, |N\rangle$ assuming
of course that we always have $F_M \neq F_N$ for $M \neq N$. We
may summarize that the main merit of the use of the QES states
lies in the compact and easily generated matrix form of our
Coulomb + harmonic matrix Schr\"{o}dinger equation at any non-QES
charge $F$.

\subsection{Outlook}

We have seen that our system (\ref{SE}) may quite efficiently be
treated by the non-numerical as well as almost purely numerical
means. In the former sense it has been shown to lie somewhere in
between the QES and exactly solvable category (cf. Table 2). In
comparison with its completely solvable neighbor (i.e., with its
own harmonic oscillator special case of ref. \cite{ptho}), its
nontrivial $b \neq 0$ and $F \neq 0$ versions do not generate {\em
all} their bound states in the elementary form. Still, in contrast
to the quasi-odd QES model \cite{I} in Table~2, its present
quasi-even partner supports an {\em infinite} number of bound
states in the closed, elementary form. Thus, our example (as well
as any one of its many analogs) could be assigned a ``midway"
status in some of its applications and interpretations.

\subsection*{Acknowledgement}

Work supported by GA AS (Czech Republic), grant Nr. A 104 8004.


\newpage

Table 1. $N-$dependence of eigencharges at $ {\cal L}=4$ and
$b=5$.

$$
 \begin{array}{||c||cccc||} \hline
\hline
 N&\multicolumn{4}{c||}{F_N}\\
 \hline
 3&
  -15.\,\allowbreak
611&  -5.\,\allowbreak927\,9 &
 4.\,\allowbreak888\,7&16.\,\allowbreak651\\
 30&
-27.\,\allowbreak
149&-9.\,\allowbreak290\,9&8.\,\allowbreak929\,4&27.\,\allowbreak511\\
 100&-44.\,\allowbreak 732&
-15 & 14.\,\allowbreak 865&44.\,\allowbreak 867\\
 200&
 -61.\,\allowbreak 665&
 -20.\,\allowbreak602&
 20.\,\allowbreak531&
 61.\,\allowbreak736\\
 300&
-74.\,\allowbreak 856&
 -24.\,\allowbreak984&24.\,\allowbreak936&74.\,\allowbreak904\\
 1000&
 -134.\,\allowbreak93&
 -44.\,\allowbreak 985&
 44.\,\allowbreak970&
 134.\,\allowbreak94\\
3000&
 -232.\,\allowbreak82&-77.\,\allowbreak 610&
 77.\,\allowbreak605&232.\,\allowbreak83\\
 30 000&
 -734.\,\allowbreak99&
 -245.\,\allowbreak00&
 245.\,\allowbreak 00&734.\,\allowbreak99\\
 \hline
 \hline
\end{array}
$$

Table 2. Solvable pseudo-Hermitian potentials: Tentative
classification.

$$
 \begin{array}{||c||c|c|c|c|c|c|c||} \hline
 \hline
 {\rm class}
 &\multicolumn{2}{c|}{\rm
 quasi-exact}
 &\multicolumn{2}{c|}{\rm intermediate }
 &\multicolumn{2}{c|}{\rm exact }
 \\
 \hline
  \multicolumn{1}{||c||}{\rm  solutions\ available }
 &
\multicolumn{2}{|c|}{{\rm at\ a\ finite\ set\ of} \  N} &
\multicolumn{2}{|c|}{{\rm at\ infinitely\ many}\ N}
   &
\multicolumn{2}{c|}{{\rm at\ all} \ N}
 \\
 \multicolumn{1}{||c||}{\rm range\ of\ couplings
  }
  & \multicolumn{2}{|c|}{\rm
 restricted}
  &
 \multicolumn{2}{|c|}{\rm restricted}
  &\multicolumn{2}{|c|}{\rm any}
 \\
 \multicolumn{1}{||c||}{\rm illustrative\ example  }
 & \multicolumn{2}{|c|}
 {\cite{I}}
  &
 \multicolumn{2}{|c|}{\rm here}
  &\multicolumn{2}{|c|}{\cite{ptho}}
 \\
   \hline
\hline
\end{array}
$$



\newpage
\begin{thebibliography}{99}


\bibitem{Hautot}
Hautot A 1972 Phys. Lett. A 38 305

\bibitem{Ushveridze}
Ushveridze A G 1994 Quasi-exactly Solvable Models in Quantum
Mechanics (Bristol: IOPP)

\bibitem{BB}
Bender C M and Boettcher S 1998 Phys. Rev. Lett. 80
 5243;

Bender C M, Boettcher S and Meisinger P N 1999 J. Math. Phys. 40
2201

\bibitem{BM}
Bender C M and Milton K A 1997 Phys. Rev. D 55 R3255 and 1998
Phys. Rev. D 57 3595

\bibitem{BBjpa}
Bender C M and Boettcher S 1998 J. Phys. A: Math. Gen. 31 L273

\bibitem{quartic}
Voros A 1983 Ann. Inst. Henri Poincar\'e 39 211;

Znojil M 2000 J. Phys. A: Math. Gen. 33 4203

\bibitem{I}
Znojil M 1999 J. Phys. A: Math. Gen. 32 4563

\bibitem{BG}
Buslaev V and Grecchi V 1993 J. Phys.  A: Math. Gen.  26 5541

\bibitem{AM}
Ahmed Z 2001 Phys. Lett. A 290 19;

Mostafazadeh A 2002 J. Math. Phys. 43 205

\bibitem{Constantinescu}
Feshbach H and Villars F 1958 Rev. Mod. Phys. 30 24

\bibitem{Ernie}
Kalnins E 2001 and Mostafazadeh A 2002, private communication

\bibitem{Geyer}
Scholtz F G, Geyer H B and Hahne F J W 1992 Ann. Phys. (NY) 213 74

\bibitem{Jolos}
Janssen D,  D\"{o}nau F,  Frauendorf S and Jolos R V 1971 Nucl.
PHys. A 172 145

\bibitem{AMb}
Mostafazadeh A 2002 J. Math. Phys. 43 2814

\bibitem{quasi}
Znojil M 2001 ``Conservation of pseudonorm in PT symmetric quantum
mechanics" (arXiv: math-ph/0104012, unpublished).


\bibitem{comment}
Znojil M 2000 Phys. Rev. A 61 066101

\bibitem{ptho}
Znojil M 1999 Phys. Lett. A 259  220


\bibitem{Quesne}
Bagchi B, Quesne C and Znojil M 2001 Mod. Phys. Lett. 16 2047

\bibitem{ptsusy}
Znojil M 2002 J. Phys. A: Math. Gen. 35 2341;

L\'{e}vai G and Znojil M 2002 arXiv: quant-ph/0206013, submitted

\bibitem{sextic}
Turbiner A 1988 Commun. Math. Phys. 118 467;

Bagchi B, Cannata F and Quesne C 2000 Phys. Lett. A 269 79;

Cannata F,  Ioffe M, Roychoudhury R and Roy P 2001 Phys. Lett. A
281 305;

Suzuki J 2001 J. Stat. Phys. 102 1029;

Dorey P, Dunning C and Tateo R 2001 J. Phys. A: Math. Gen. 34 5679
and L391

\bibitem{decadic}
Znojil M 2000 J. Phys. A: Math. Gen. 33 6825


\end{thebibliography}
\end{document}